\documentclass[useAMS,usenatbib]{mn2e}

\usepackage{graphicx} 

\title[UFOs in the WAX sample]{On the presence of ultra-fast outflows in the WAX sample of Seyfert galaxies}
\author[F. Tombesi et al.]{F. Tombesi$^{1,2}$\thanks{E-mail: ftombesi@astro.umd.edu} and M. Cappi$^{3}$\\
$^{1}$X-ray Astrophysics Laboratory, NASA/Goddard Space Flight Center, Greenbelt, MD 20771, USA\\
$^{2}$Department of Astronomy, University of Maryland, College Park, MD 20742, USA\\
$^3$INAF-IASF Bologna, Via Gobetti 101, I-40129 Bologna, Italy}
\begin{document}

\date{Accepted ???. Received ???; in original form ???}


\maketitle

\label{firstpage}

\begin{abstract}
The study of winds in active galactic nuclei (AGN) is of utmost
importance as they may provide the long sought-after link between the central
black hole and the host galaxy, establishing the AGN
feedback. Recently, Laha et al.~(2014) reported the X-ray analysis of a
sample of 26 Seyferts observed with \emph{XMM-Newton}, which are part
of the so-called warm absorbers in X-rays (WAX) sample. 
They claim the non-detection of Fe K absorbers indicative of
ultra-fast outflows (UFOs) in four observations previously analyzed by
Tombesi et al.~(2010a). They mainly impute the Tombesi et al. detections to an
improper modeling of the underlying continuum in the E$=$4--10~keV band.
We therefore re-address here the robustness of
these detections and we find that the main reason for the claimed
non-detections is likely due to their use of single events only spectra, which
reduces the total counts by 40\%. Performing a re-analysis of the data
in the whole E$=$0.3--10~keV energy band using their models and spectra including also double events, we find 
that the blue-shifted Fe K absorption lines are indeed detected at $>$99\%. This
work demonstrates the robustness of these detections in
\emph{XMM-Newton} even including
complex model components such as reflection, relativistic lines and warm
absorbers. 
\end{abstract}


\section{Introduction}

Mounting evidences for the presence of highly ionized, high velocity
outflows along the line of sight of bright active galactic nuclei
(AGNs) have been obtained since the last decade. This material shows up
preferentially in the X-ray spectra as blue-shifted K-shell absorption
lines from Fe~XXV and Fe~XXVI at energies E$\ga$7~keV of both
high/low-z radio-quiet and radio-loud sources (e.g., 
Pounds et al.~2003a; Dadina et al.~2005; Markowitz, Reeves \& Braito
2006; Braito et al.~2007; Chartas et al. 2009; Cappi et al.~2009; Reeves
et al.~2014; Tombesi et al.~2010a, 2010b, 2011a, 2011b, 2012b, 2013b;
Ballo et al.~2011; Giustini et al.~2011; Patrick et al.~2012; Pounds
\& Vaughan 2012; Gofford et al.~2013). 

Often the blue-shifts of these lines imply an outflow velocity higher than 10,000~km~s$^{-1}$, in which cases the absorbers are indicated as ultra-fast outflows (UFOs). This velocity threshold is arbitrary, being chosen only to initially differentiate with the slower ($v_{out}$$\simeq$100--1,000~km~s$^{-1}$), less ionized warm absorbers (WAs) commonly detected in the soft X-rays. The location and kinetic power of the UFOs indicate that they could represent powerful accretion disk winds, capable to exert a significant feedback on the surrounding environment (e.g., King \& Pounds 2003; Tombesi et al.~2012a). Possible links between the UFOs and the WAs have been recently explored (Fukumura et al.~2010, 2014; Pounds \& Vaughan 2012; Tombesi et al.~2013a; King \& Pounds 2014).  
Recently Laha et al.~(2014) reported the analysis of a sample of 26
Seyferts 1s observed with \emph{XMM-Newton}, the so-called warm
absorbers in X-rays (WAX) sample. The main objective of that
work was to estimate the parameters of the WAs in these sources
performing a combined RGS and EPIC-pn data analysis in the
E$=$0.3--10~keV band. Laha et al.~(2014) also performed a eye
inspection of the residuals of their models in the E$=$6--8.5~keV
energy band for six sources with UFOs reported in Tombesi et
al.~(2010a) (see their Fig.~A6) and, combined with other evidences
from the literature, claimed that Fe K absorption lines are not
present in these cases. Here, we consider a critical analysis of these
statements for these six sources, namely MRK~509, ARK~120, UGC~3973,
NGC~4051, MRK~766 and IC~4329A, and demonstrate the presence of UFOs
in these datasets and discuss the reasons why they reached different conclusions.

\section[]{Data analysis and results}

The \emph{XMM-Newton} EPIC-pn spectra were extracted following the
method of Laha et al.~(2014).  We use the latest calibration files and
software packages \verb+heasoft+ v.~6.15, \verb+sas+
v.~13.5 and \verb+XSPEC+ v.~12.8.1 (as of April 2014).
Here, the inspection of the Fe K band region is performed using both ratios
 and energy-intensity contour plots, as described in \S~3.2 of Tombesi
 et al.~(2010a). The neutral Fe K$\alpha$ emission line and related
 reflection component are modeled with \emph{pexmon} in
 \verb+XSPEC+, as Laha et al.~(2014).  When indicated by Laha et
 al.~(2014), relativistic broad lines are modeled with \emph{laor} or
 \emph{diskline} in \verb-XSPEC- assuming an outer radius of
 $r_{out}$$=$400$r_g$ and an inner radius of $r_{in}$$=$1.235~$r_g$
 and $r_{in}$$=$10~$r_g$, respectively. The WAs are modeled with an
 XSTAR table with turbulent velocity of 100~km~s$^{-1}$
 (as Laha et al.~2014) and an ionizing continuum with a slope
 $\Gamma$$\simeq$2, consistent with the average for Seyfert 1s (Tombesi et al.~2011a). 
Throughout, we consider the 1$\sigma$ errors for one additional parameter, if not otherwise stated.

We note that Laha et al.~(2014) extracted the spectra using only
single events because this provides a slightly better energy
resolution and they represents the bulk of events at E$<$2~keV. 
While this choice could be seen as appropriate for their soft X-ray focused
analysis, it causes a significant loss of counts with respect to
considering both singles and doubles, about  40\% less in the
4--10~keV band\footnote{http://xmm2.esac.esa.int/docs/documents/CAL-TN-0018.pdf}. 
As shown below, this is in our opinion the main reason why Laha et al.~(2014) did not report
the detection of any Fe K absorption lines. We find that the
lines are indeed present in the single events spectra with parameters
consistent within the 1$\sigma$ uncertainties. However, given the much lower
signal-to-noise, their significance is lower
than 99\% in three out of four cases, namely MRK~509 (96\%), ARK~120
(95\%) and IC~4329A (98.5\%). 

The EPIC-pn data in Laha et al.~(2014) are grouped to a minimum of 20 counts
per energy bin and at most 5 energy bins per resolution element. We
find this probably not to be the best choice when fitting the wide 0.3--10~keV band
given the large difference in counts between low and high energies. 
In fact, the binning fixed to the energy resolution will give a higher
statistical weight at those energy points with higher number of counts per
bin, thus even small fluctuations in the soft X-rays can inflate the
$\chi^2$ distribution. 
This effect is reduced allowing a more similar statistical weight for
each data point, for instance just rebinning to a minimum of 25
counts per bin. 
The same best-fit model can result in a higher $\chi^2/\nu$ in the
former case, as seen in several values reported in Table~5 of Laha et al.~(2014). 
Moreover, given that the F-test probability depends on  $\chi^2/\nu$,
for the same best-fit model and the same value of
$\Delta\chi^2/\Delta\nu$, the resultant probabily is systematically
lower in the former case. This effect is negligible
considering a limited energy band, e.g., E$=$4--10~keV (Tombesi et
al.~2011a).

In the following, we will test the possible model dependence of the Fe
K absorption lines using the EPIC-pn spectra (including both single and double
events) and rebinning the data to a minimum of
25 counts per bin.

\subsection{MRK~509}

Laha et al.~(2014) reported the combined RGS and EPIC-pn analysis of
one \emph{XMM-Newton} observation of MRK~509 performed in October 2005
(OBSID 0306090201). They claim that there are no Fe K absorption features
in the 6--8.5~keV band after an eye-inspection of the
residuals in their Fig.~A6. This is not consistent with the detection of an absorption
line at the observed energy of E$=$7.76~keV reported by Cappi
et al.~(2009) and Tombesi et al.~(2010a). We note that
absorption residuals of $\sim$15\% of the continuum are indeed present in Laha et al.~(2014). Here we use their broad-band model to fit the broad-band EPIC-pn spectrum and perform a more sophisticated analysis using the energy-intensity
contour plots. 

In Table~5, Laha et al.~(2014) indicate a power-law continuum with
$\Gamma$$=$$2.13$ and two black-body components to model the soft excess with temperatures
$kT$$=$79~eV and $kT$$=$161~eV, respectively. They also require a
high reflection fraction of $R=2.3$. We assume the
same parameters for the soft X-ray emission lines and warm
absorbers. From their Table~6 we have emission lines at $E$$=$0.562~keV and
$E$$=$0.598~keV and a warm absorber with
log$\xi$$=$3.24~erg~s$^{-1}$~cm$^{-2}$, log$N_{H}$$=$20.8~cm$^{-2}$
and outflow velocity $v_{out}$$=$$-6500$~km~s$^{-1}$.
They also include a broad relativistic line modeled with a \emph{laor} profile. 
Thus, they assume a maximally spinning black hole,
a high emissivity profile
$\beta$$=$4.53 and an inclination $i$$=$48$^{\circ}$. The reported
observed energy and high equivalent width are E$=$6.63~keV and
EW$=$310~eV, respectively.

 We note that such steep power-law continuum, high reflection
 fraction and extreme relativistic line have never been reported before for
 this source (e.g., Dadina 2007; Patrick et al.~2012; Gofford et
 al.~2013; Nandra et al.~2007). Also, the soft X-rays of this and other sources could be complicated by the
 presence of thermal Comptonization components (e.g., Petrucci et
 al.~2013). In fact, we find that this model provides a sufficient fit
 to the data only with an extremely low \emph{pexmon} iron
abundance of $A_{Fe}$$=$$0.09\pm0.01$. Moreover, as already indicated
by Ponti et al.~(2013), the residual emission at E$\simeq$6.4~keV in
the contour plots in Fig.~1 can be accounted for by an additional emission line with
$\sigma$$\simeq$100~eV produced in the broad line region.

As it can be clearly seen in the data-to-model ratios and
energy-intensity contour plots with respect to this broad-band model in Fig.~1, an
absorption line at the observed energy of E$_{obs}$$\simeq$7.75~keV equivalent to that reported by Cappi et al.~(2009)
and Tombesi et al.~(2010a) is indeed still present. The line parameters
are E$=$$8.04^{+0.03}_{-0.02}$~keV  (rest frame), $\sigma$$=$0 (unresolved),
$I$$=$$(-5.4\pm1.5)\times 10^{-6}$~ph~s$^{-1}$~cm$^{-2}$ and
EW$=$$-20\pm6$~eV. The line is required at a level of 99.6\%
($\Delta\chi^2/\Delta\nu = 12.3/2$). The fit statistics is
$\chi^2/\nu = 2008.8/1796$.

We note absorption lines at E$>$7~keV
in MRK~509 were initially reported by Dadina et al.~(2005) in
\emph{XMM-Newton} and \emph{BeppoSAX} observations performed in 2000--2001. Cappi et
al.~(2009) reported detections in 3/5 \emph{XMM-Newton} observations
obtained in 2000--2006. These were confirmed by the re-analysis of
Tombesi et al.~(2010a). However, we note that in a 2009 \emph{XMM-Newton}
campaign it was possible only to place upper limits on the presence of
highly ionized Fe K absorption (Ponti et al.~2013).

\subsection{ARK~120}

Laha et al.~(2014) analyzed the spectrum of one \emph{XMM-Newton} observation of ARK~120
(OBSID 0147190101). Tombesi et al.~(2010a) reported the detection of a
blue-shifted Fe K absorption line at the observed energy of
E$=$$8.89\pm0.03$~keV. In their eye-inspection of
the Fe K band in Fig.~A6, Laha et al.~(2014) claim the non-detection of such
feature. However, we note that this might simply be due to the fact that the
feature is outside of the energy band considered of E$=$6--8.5~keV.

Anyway, the broad-band model assumed by Laha et al.~(2014) consists of a
power-law continuum with $\Gamma$$=$1.97, two black-body components to
model the soft excess with temperatures $kT$$=$102~eV and
$kT$$=$240~eV (see their Table~5). They also include a soft X-ray
emission line at the energy E$=$0.554~keV and a \emph{pexmon}
reflection component with $R$$=$0.75. They do not report a significant detection
of a warm absorber in this source (see their Table~6). Finally, they
also include a putative \emph{diskline} with parameters E$=$6.9~keV,
$\beta$$=$$-3.8$, $i$$<$50$^{\circ}$. In this case they implicitly assume a non-spinning black hole.
We find again that this model can fit the data only for a very low
\emph{pexmon} iron abundance of $A_{Fe}$$=$$0.40^{+0.02}_{-0.04}$ and a low
inclination of the \emph{diskline} of $i$$\simeq$13$^{\circ}$. 

As we can see for the ratios and the contour plots in Fig.~1, the line at the observed energy of E$=$8.89~keV is
indeed present. It has an intensity corresponding to about 20\% of the continuum.
The line parameters are E$=$$9.18\pm0.03$~keV  (rest frame), $\sigma$$=$0 (unresolved),
$I$$=$$(-6.5\pm1.7)\times 10^{-6}$~ph~s$^{-1}$~cm$^{-2}$,
EW$=$$-31^{+7}_{-9}$~eV. These values are consistent with those
reported by Tombesi et al.~(2010a) in their Table~A.2. The detection
confidence level is 99.3\% ($\Delta\chi^2/\Delta\nu = 13.8/2$). The fit statistics
is $\chi^2/\nu = 2449.6/1776$.

 \begin{figure*}
 \centering
  \includegraphics[width=5.5cm,height=5.5cm,angle=0]{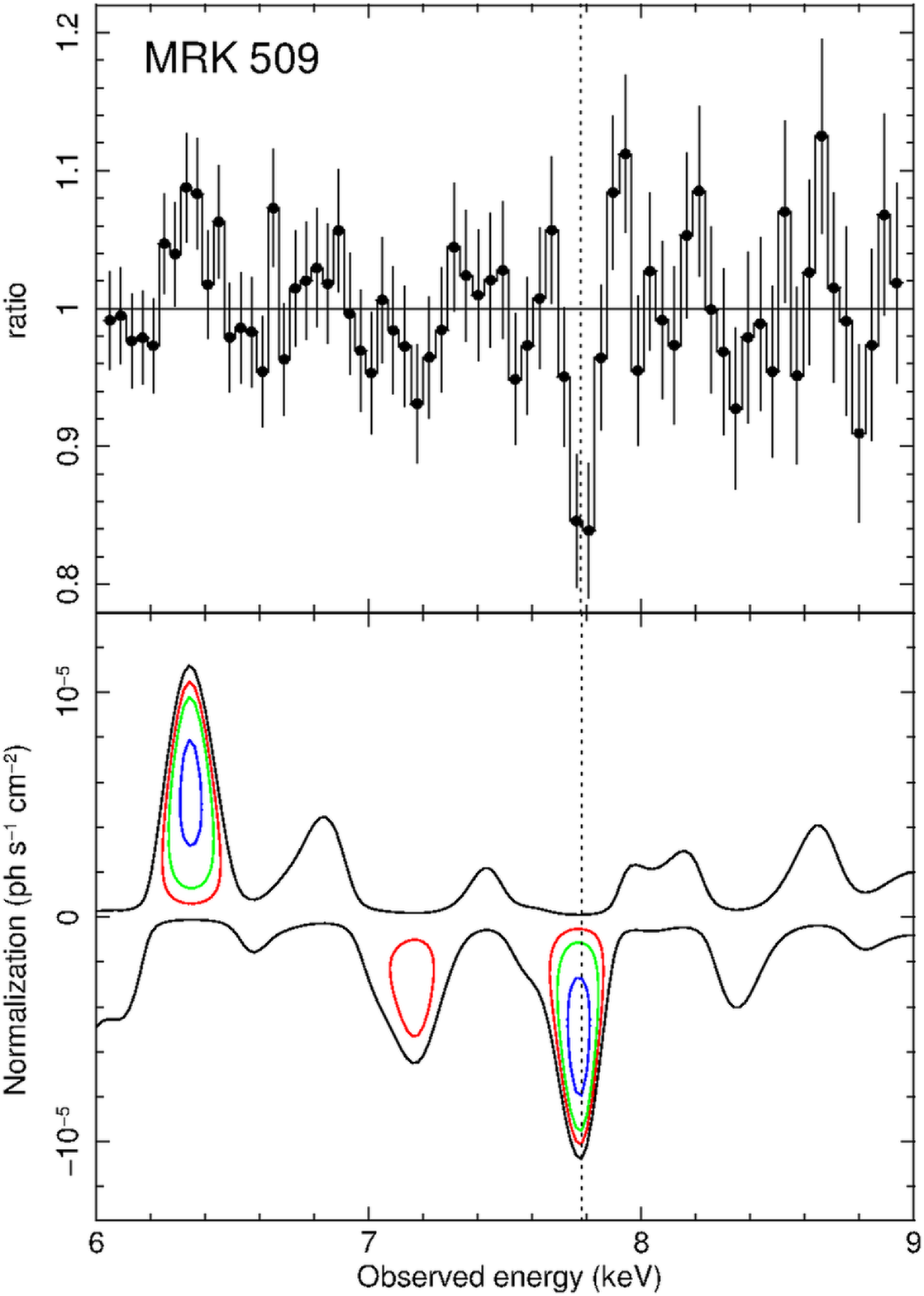}
\hspace{1cm}
  \includegraphics[width=5.5cm,height=5.5cm,angle=0]{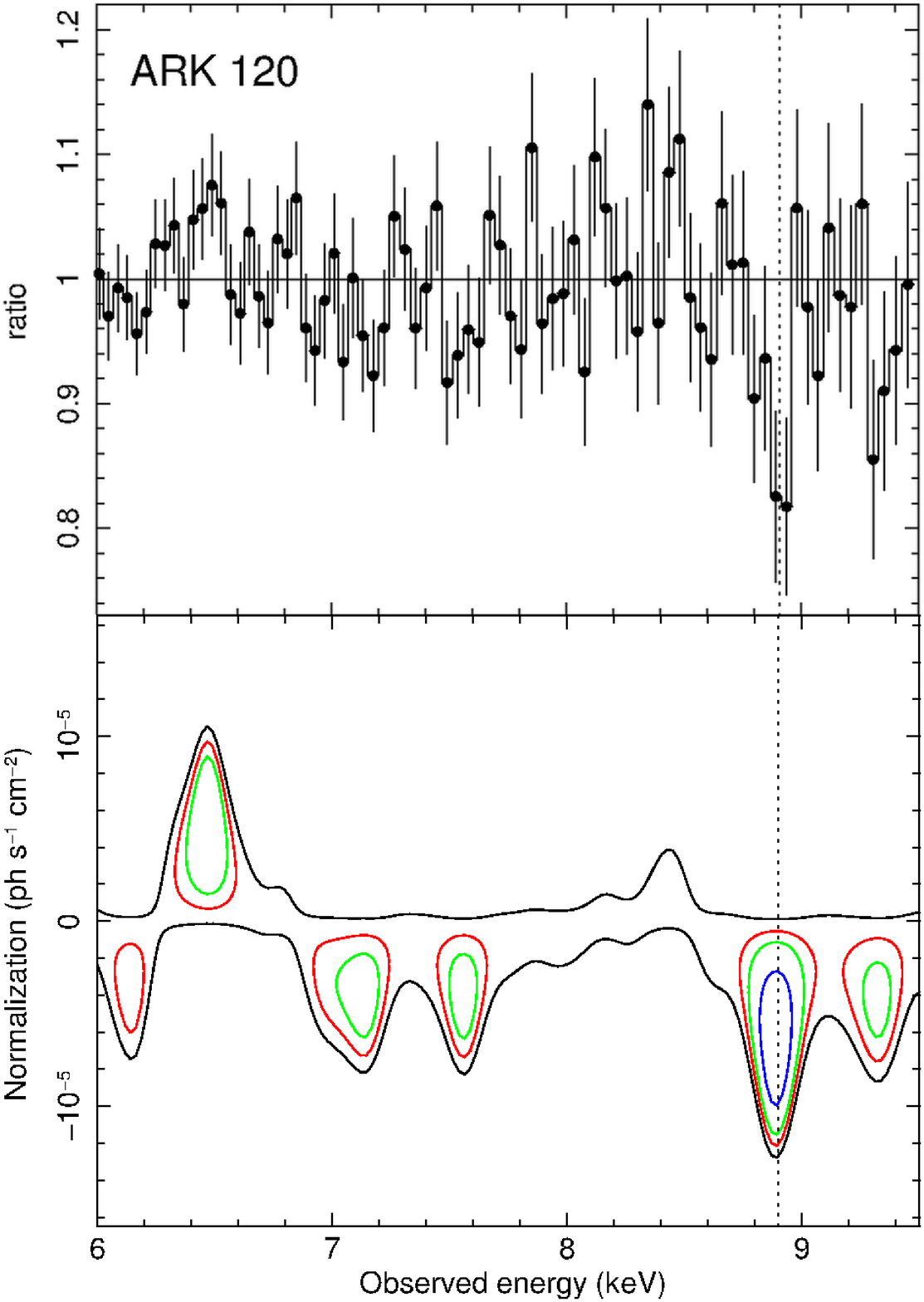}
  \includegraphics[width=5.5cm,height=5.5cm,angle=0]{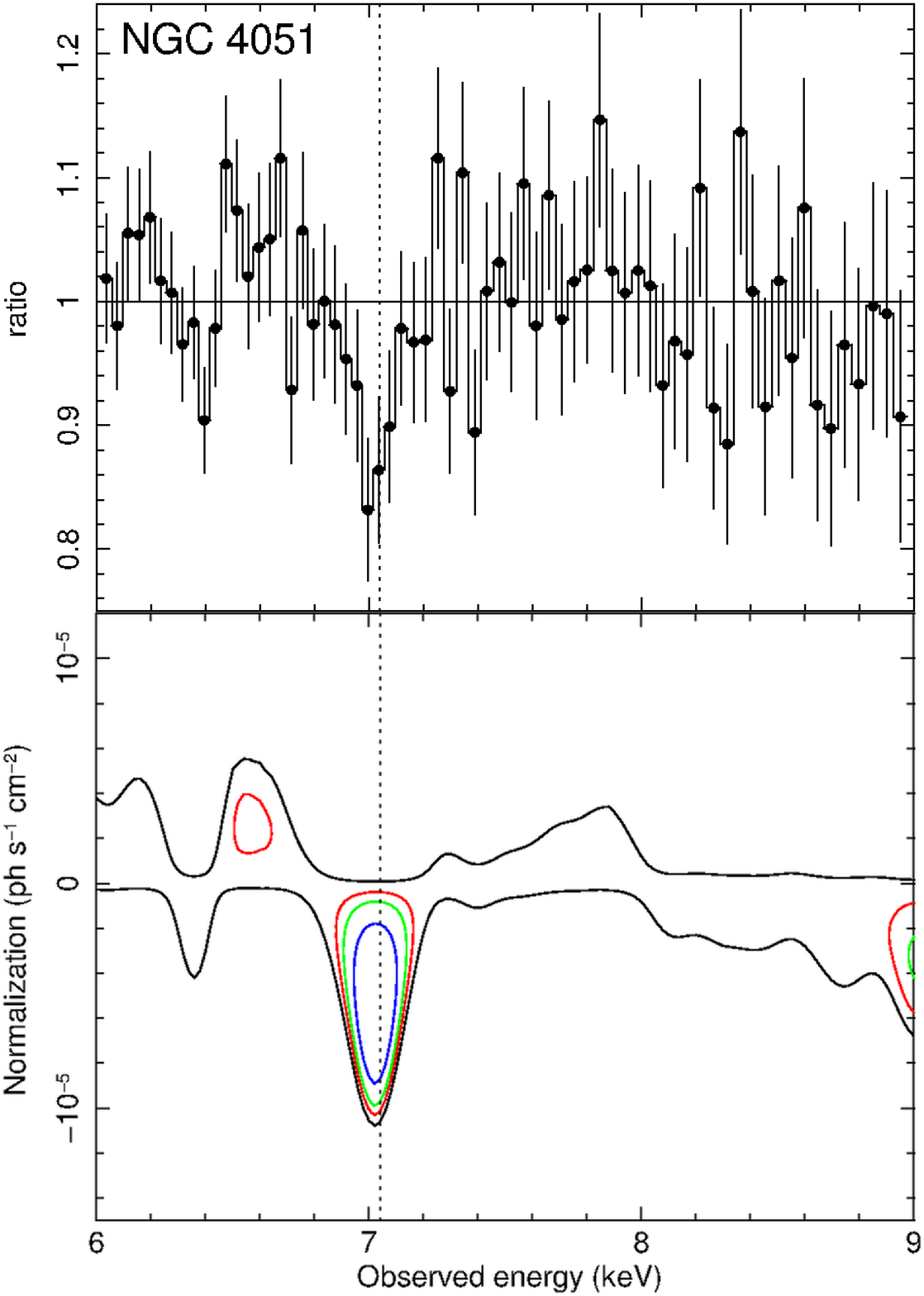}
\hspace{1cm}
  \includegraphics[width=5.5cm,height=5.5cm,angle=0]{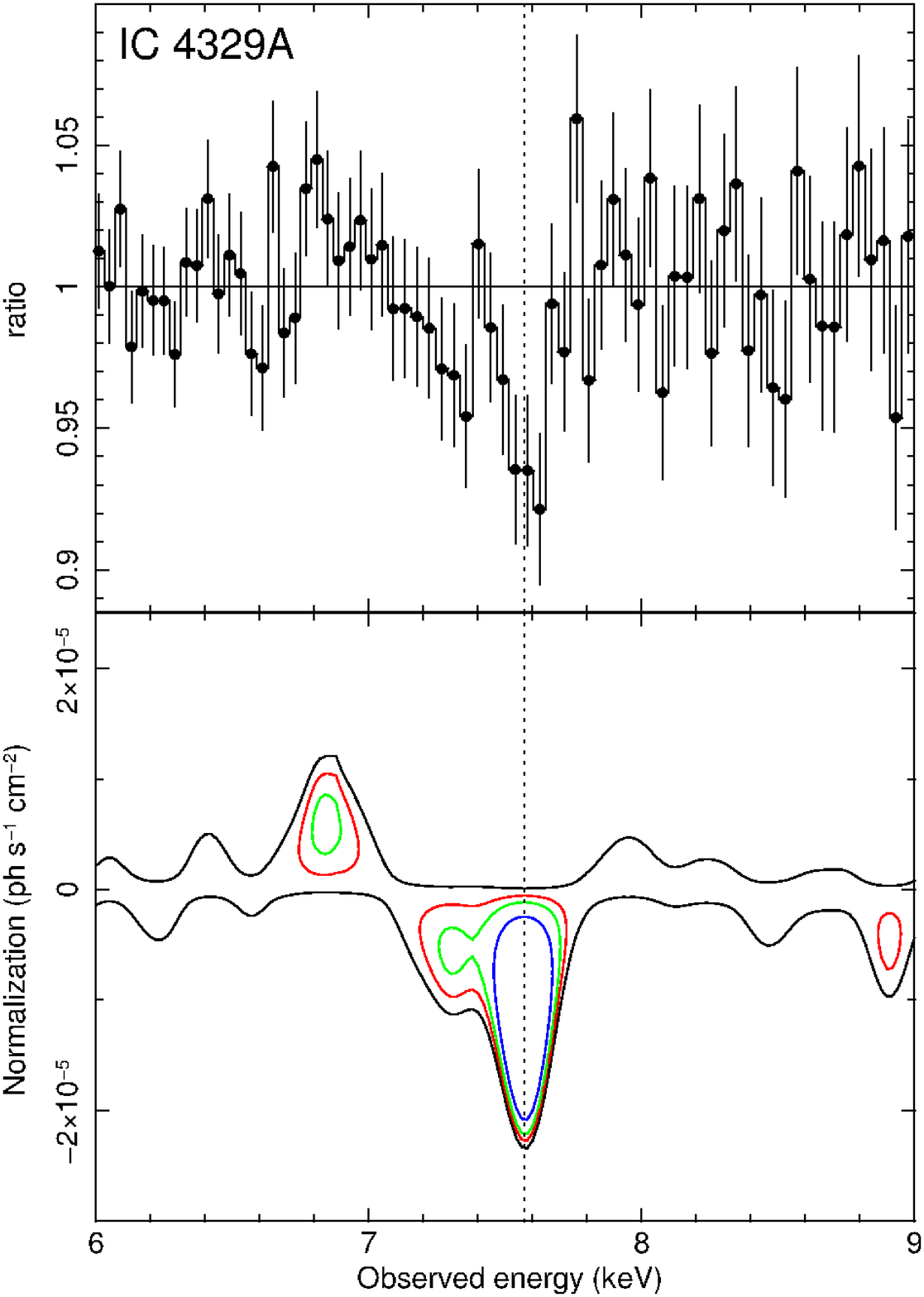}
   \caption{The zoomed fits in the Fe K band of the EPIC-pn spectra of
     the sources in the WAX sample of Laha et al.~(2014) and
     previously analyzed by Tombesi et al.~(2010a). Broad-band models
     equivalent to those of Laha et al.~(2014) are fitted in the
     E$=$0.3--10~keV band. The upper panels show the ratios between
     the spectra and the models. The data are rebinned to 1/5 of the
     FWHM of the instrument for clarity.  The lower panels show the
     energy-intensity contour plots, with the contours referring to
     the confidence levels of 68\% (red), 90\% (green) and 99\%
     (blue), respectively. The vertical dotted lines indicate the
     location of the detected blue-shifted Fe K absorption lines.}
   \end{figure*}

\subsection{UGC 3973}

Laha et al.~(2014) considered one \emph{XMM-Newton} observation of UGC~3973 (or MRK~79)
obtained in April 2008 (OBSID 0502091001). This observation was not in the sample of Tombesi et
al.~(2010a), who considered three observations obtained between 2006 and 2007. 
Tombesi et al.~(2010a, 2011a) reported the detection of a blue-shifted Fe K
absorption line at E$=$7.63~keV indicative of a UFOs with velocity
$v_\mathrm{out}$$\simeq$0.09c in the first observation (OBSID
0400070201). The fact that Laha et al.~(2014) did not observe an intense Fe K
absorption line in their spectrum could be due to variability
of the absorber and/or to the lower statistics of this fainter
observation (also limited by the use of single events only).

\subsection{NGC~4051}

Blue-shifted Fe K absorption lines have been
reported by many authors in this source using both \emph{XMM-Newton} and
\emph{Suzaku}. Tombesi et al.~(2010a, 2011a) analyzed the first two \emph{XMM-Newton} observations of this source 
(OBSID 0109141401 and 0157560101) and reported highly
ionized Fe K outflows with velocities of
$v_{out}$$\simeq$11,000~km~s$^{-1}$ and $v_{out}$$\simeq$0.2c,
respectively. We note that Fe K outflows with velocities of $\sim$6,000~km~s$^{-1}$
and $\sim$0.1c were reported also by several other authors using
\emph{Chandra}, \emph{XMM-Newton} and \emph{Suzaku} data (Lobban et
al.~2011; Patrick et al.~2012; Pounds \& Vaughan 2012; Gofford et al.~2013).

Laha et al.~(2014) considered the first \emph{XMM-Newton} observation
of NGC~4051 (OBSID 0109141401) in their WAX sample. From an eye
inspection of the E$=$6--8.5~keV residuals in their Fig.~A6, they claim that
the absorption line at the observed energy of E$\simeq$7.06~keV
reported by Tombesi et al.~(2010a) is not
detected. In order to check whether this was due to their broad-band modeling or
to an inadequate analysis of the Fe K band data, we re-analyze the EPIC-pn
spectrum using their model. We note that this feature is detectable
also in the single events spectrum at $>$99\%. 

They consider a rather steep power-law continuum with $\Gamma$$=$2.35, a
black-body component with $kT$$=$101~eV to model the soft excess and a
\emph{pexmon} neutral reflection component with a very high reflection
fraction of $R$$\simeq$5 (see their Table~5). In the soft X-rays they
consider two emission lines at E$=$0.561~keV and E$=$0.597~keV. We include two warm absorbers. The first with log$\xi$$=$0.28~erg~s$^{-1}$~cm,
log$N_H$$=$20.43~cm$^{-2}$ and $v_{out}$$=$$-600$~km~s$^{-1}$. The second with log$\xi$$=$2.87~erg~s$^{-1}$~cm,
log$N_H$$=$22.39~cm$^{-2}$ and $v_{out}$$=$$-688$~km~s$^{-1}$ (see
their Table~6).

Finally, they also include a putative broad disk line using the
\emph{laor} model. They require extreme parameters, with a
line energy of E$=$7~keV, large equivalent width of EW$=$461~eV, high
emissivity profile of 7.6, and inclination of $i$$<$41$^{\circ}$ (see
their Table~7). The use of \emph{laor}, which assumes a maximally
spinning black hole, is not justified. Moreover, such an extreme broad
line has never been reported before (e.g., Nandra et al.~2007; Patrick et al.~2012). 

We find that in order to provide a sufficient representation of the
spectrum this model requires again a very low \emph{pexmon} iron abundance of $A_{Fe}$$=$$0.19\pm0.01$. 
From the spectral ratios and the contour plots in Fig.~1, we note that
an absorption feature at E$\simeq$7.06~keV is indeed present in the
data. In fact, the
inclusion of an absorption line gives the same parameters as those
reported in Table~A.2 of Tombesi et al.~(2010a), i.e.,
E$=$$7.04^{+0.03}_{-0.02}$~keV (rest frame), $\sigma=0$~eV (unresolved),
$I$$=$$(-5.3\pm1.3)\times 10^{-6}$~ph~s$^{-1}$~cm$^{-2}$ and 
EW$=$$-28\pm7$~eV. The line is required at 99.8\% ($\Delta\chi^2/\Delta\nu = 15/2$). The fit statistics is $\chi^2/\nu = 1857.2/1593$.

\subsection{MRK~766}

This source is well known to show complex and highly variable absorption in the Fe K band due to outflowing winds with
velocities of the order of $v_{out}$$\simeq$10,000--20,000~km~s$^{-1}$ from
\emph{XMM-Newton} and \emph{Suzaku} data (e.g., Pounds et al.~2003b; Miller et al.~2007;
Turner et al~2007; Risaliti et al.~2011; Patrick et al.~2012; Gofford et al.~2013). 

Laha et al.~(2014) report the non detection of Fe K absorption lines
in an \emph{XMM-Newton} observation performed in 2001 (OBSID
0109141301). We note that Tombesi et al.~(2010a, 2011a) also did no report a
detection in this case. However, Tombesi et al.~(2010a) analyzed 8
\emph{XMM-Newton} observations of MRK~766 and reported Fe K absorption
lines indicative of UFOs with a velocity of $v_{out}$$\simeq$20,000~km~s$^{-1}$ in two
cases (OBSID 0304030301 and 0304030501). Given the high X-ray flux variability and the 
highly ionized absorbers in this source, a thorough inspection of the Fe K band would require a time-resolved spectral analysis (e.g., Miller et al.~2007; Turner et al~2007; Risaliti et al.~2011).

\subsection{IC~4329A}

Laha et al.~(2014) report the analysis of the \emph{XMM-Newton}
observation of IC~4329A (OBSID 0147440101) performed in 2003. They
claim the non-detection of an absorption line at the observed energy of
E$=$$7.57\pm0.03$~keV previously reported by Tombesi et al.~(2010a).  
We note that this absorption line was initially reported by Markowitz, Reeves \& Braito (2006), who performed a detailed
broad-band analysis of the same \emph{XMM-Newton} dataset.


We re-analyze the data applying Laha et al.~(2014) model. 
We consider a neutral absorbed ($N_H$$=$$3.8\times 10^{21}$~cm$^{-2}$)
power-law continuum with $\Gamma$$\simeq$1.8 and two black-body components
with temperatures $kT$$=$46~eV and $kT$$=$286~eV to model the soft
excess (see their Table~5). 
We consider also a \emph{pexmon} component with $R$$=$1.67 and an
emission line at E$=$6.87~keV. We include two soft X-ray emission lines
at E$=$0.528~keV and E$=$0.649~keV and three warm absorbers. The first
with log$\xi$$=$$-0.58$~erg~s$^{-1}$~cm, log$N_H$$=$20.96~cm$^{-2}$
and $v_{out}$$=$$-1020$~km~s$^{-1}$. The second with
log$\xi$$=$$1.87$~erg~s$^{-1}$~cm, log$N_H$$=$20.54~cm$^{-2}$ and
$v_{out}$$=$$-660$~km~s$^{-1}$. The third with
log$\xi$$=$$3.33$~erg~s$^{-1}$~cm, log$N_H$$=$21.27~cm$^{-2}$ and
$v_{out}$$=$$-990$~km~s$^{-1}$ (see their Table~5 and Table~6).
Finally, we include also a putative broad line, parameterized with
\emph{diskline}, with E$=$6.3~keV, $\beta$$=$$-2.19$ and
$i$$=$30$^{\circ}$. This implicitly assumes a non-spinning black hole. 
The fit requires a \emph{pexmon} iron abundance of $A_{Fe}$$=$$0.95^{+0.25}_{-0.50}$.

From the ratios and the contour plots in Fig.~1 we note the presence of an absorption feature at the
observed energy of E$\simeq$7.57~keV. Indeed, this absorption line with rest-frame
energy E$=$$7.70^{+0.02}_{-0.03}$~keV, intensity $I$$=$$(-11.8\pm2.5)\times
10^{-6}$~ph~s$^{-1}$~cm$^{-2}$ and EW$=$$-15^{+4}_{-3}$~eV is required
at 99.99\% ($\Delta\chi^2/\Delta\nu = 20.4/2$). The fit statistics is $\chi^2/\nu = 2200.3/1925$.
The absorption line parameters are equivalent to those reported by
Markowitz, Reeves \& Braito (2006) and Tombesi et al.~(2010a).

\section[]{Discussion and conclusions}

We report a critical examination of the claims of Laha et
al.~(2014) regarding the absence of UFO detections in six \emph{XMM-Newton}
EPIC-pn observations of six sources of their WAX sample. Four
of these were previously analyzed by Tombesi et
al.~(2010a). The non-detection is mainly imputed to an improper modeling of
the E$=$4--10~keV band. Here, we show a re-analysis of the data following their procedure
and using their models in the whole E$=$0.3--10~keV interval. 

The main reason why Laha et al.~(2014) did not find any Fe K
absorption lines is likely due to their selection of only single
events spectra, thus losing 40\% of the total counts. In fact, we find
that the lines are present in the single events spectra, although
with a significance slightly lower than 99\% in 3 out of 4 cases. In
NGC~4051 the line is present at $>$99\% also in the
single events spectrum.   
Using the single and double events spectra we find that the
absorption lines are indeed present at $>$99\% in the four cases
considered (MRK~509, NGC~4051, 
IC~4329A and ARK~120), with parameters equivalent to
those previously reported by Tombesi et al.~(2010a). 
Some of these lines might have also been missed in the eye-inspection
of the residuals in Fig.~A6 by Laha et al.~(2014). 

Most of the soft X-ray WAs in Laha et al.~(2014) have low column
    densities ($N_H$$\la$$10^{21}$~cm$^{-1}$), implying a very limited
    influence in the Fe K band. In fact, we checked that the Fe K
    absorption lines are present both with and without their inclusion. 

Overall, this re-analysis shows the robustness of the detection of the Fe K
absorption lines in the broad-band \emph{XMM-Newton} E$=$0.3--10~keV
spectra using complex model components such as reflection,
relativistic lines and warm absorbers. This is consistent with the more detailed
\emph{Suzaku} and \emph{Swift} BAT 0.6--100~keV spectral
analyses of Patrick et al.~(2012) and
Gofford et al.~(2013).

We warn that although the models assumed by Laha et al.~(2014) provide an overall good
representation of the data, they might be affected by systematics. For
instance, their models often require a steep
power-law, very high neutral reflection
fraction, very low iron abundance and very broad disk lines. Many of these parameters
are not consistent with previous studies (e.g., Dadina 2007; Nandra et
al.~2007; Patrick et al.~2012; Pounds \& Vaughan 2012; Gofford et
al.~2013). Moreover, a thorough constraint of the continuum and
reflection requires the use of instruments with sensitivity above
10~keV (e.g., Patrick et al.~2012; Gofford et al.~2013). 

A better modeling and characterization of the Fe K
absorption lines will be provided by the simultaneous use of the
calorimeter and broad-band coverage 
offered by ASTRO-H (Takahashi et al.~2012).

\section*{Acknowledgments}

FT and MC would like to thank M. Guainazzi for the comments on the analysis.
FT would like to thank T. Yaqoob and C.~S. Reynolds for the useful discussions.
MC would like to thank M. Dadina for the useful comments.

\end{document}